\input{aipcheck}

\documentclass[final]{aipproc}
\usepackage{latexsym,amssymb,pifont,amsmath,xspace}
\layoutstyle{6x9}

\begin{document}

\title{Introduction to Quantum Mechanics}


\author{Eduardo J. S. Villase\~nor}{address={Grupo de
Modelizaci\'on y Simulaci\'on Num\'erica, Universidad Carlos III
de Madrid, Avda. de la Universidad 30, 28911 Legan\'es, Spain}}

\begin{abstract}
The purpose of this contribution is to give a very brief introduction to Quantum Mechanics for an audience of mathematicians. I will follow Segal's approach to Quantum Mechanics paying special attention to algebraic issues. The usual representation of Quantum Mechanics on Hilbert spaces is also discussed.
\end{abstract}

\maketitle


\section{Introduction}

Quantum Mechanics is usually thought of as the physical theory  with the broadest range of applicability. Our current believe is that, in principle, it can be applied to any physical system, from the subatomic world to the whole universe. However, the scales at which quantum mechanical effects are significant are usually small because they are governed by one of the smallest constants in physics: the (reduced) Planck constant $\hbar\sim  10^{-34} J\cdot s$. The extremely small  value of $\hbar$ essentially excludes quantum physics from our everyday experience. Quantum Mechanics (QM) takes his name from the fact that certain physical observables were experimentally seen to take only discrete (quantized) values contrary to the theoretical predictions of Classical Mechanics. Historically, the first evidence for the discreteness of  atomic energies was provided by Franck and Hertz in 1914. Nowadays this discrete character is  very well established  experimentally because current technology allows us to resolve the atomic energy levels in exquisite detail.

The mathematical framework of QM was firmly founded by J. von Neumann in the early 1930s. The basic rules of the so called Copenhagen interpretation of QM could be summarized as follows: Any physical observable is represented in QM by a self-adjoint operator $\hat{A}$ on a Hilbert space $\mathfrak{H}$. A (pure) state of a system is defined in terms of an equivalence class of unit vectors in $\mathfrak{H}$, where two unit vectors $\psi_1$ and $\psi_2$ are equivalent if $\psi_1=z \psi_2$ for some $z\in \mathbb{C}$ such that $|z|=1$. The expectation  value of $\hat{A}$ in the state $\psi$ is given in terms of the scalar product of $\mathfrak{H}$ by $\langle \psi \,|\, \hat{A}\psi\rangle$. Finally,  the dynamical evolution of the system is determined by the specification of a self-adjoint operator $\hat{H}$ through one of the following rules: $\psi\mapsto \psi_t=\exp( it\hat{H})\psi$ or
$\hat{A} \mapsto \hat{A}_t = \exp(it\hat{H})\hat{A}\exp(- it\hat{H})$.

There is a huge bibliography on the mathematical foundations of QM. I will only mention two  excellent classical texts. First of all, the masterpiece \cite{vN:1965}, written in 1932 by J. von Neumann, is a must read. Here, von Neumann gave  the first complete  mathematical formulation of the Copenhagen interpretation of QM. Another notable reference is the one of G. Mackey  \cite{Mackey}. This book contains the seminal ideas that gave rise to the  nowadays well established  area of Quantum Logic.  Within the recent literature there are several books that a mathematician can use to approach this subject. A very interesting one, suitable for undergraduates due to its introductory character, is \cite{Strocchi}. Strocchi's book offers an excellent and short presentation of the mathematical aspects of QM using the $C^*$-algebraic structure of the set of observables that defines a physical system. The $C^*$ approach is the appropriate one to deal with field theories such as General Relativity. Another elegant and clear introduction to QM, that discusses in deep detail the fundamental conceptual problems of quantum theory, is the book of C. Isham \cite{Isham}. Then, if you really want to learn QM, you should just stop reading these notes now and get some (or all!) of those books in your favorite bookstore. If you are still reading, what you will find in the following is a kind of taster menu of the  mathematical ideas behind QM. This was actually the purpose of the talk on which these notes are based.

\section{Physical systems}
The `things' of the physical world in which  physical theories are  interested in are called physical systems or simply systems. At first sight, it is not difficult to write down many examples of them. For instance, a free point particle, a particle constrained to move on a smooth surface, or electric and magnetic fields without sources (all of them in Euclidean space) are usually referred to as systems in the physical literature. However, sooner or later,  we must give a precise definition of what we mean by all these very loosely defined concepts. At that point, in order to avoid metaphysical temptations, it is very convenient to adopt an operational point of view. In this approach pioneered by Segal \cite{Segal1}, a system is defined by the class of physical properties that can be measured on it by using concrete physical devices.  A system is then defined by its family $\mathcal{O}$ of observables. Of course, in any mathematical model of a system, the set $\mathcal{O}$ will be endowed with certain algebraic and metric properties. Given a physical system described by  $\mathcal{O}$, we can consider also the set  $\mathcal{S}$ of its states. States are characterized by the results of the measurements of all the observables in the following sense: Given a state $\omega \in \mathcal{S}$, for any $A\in\mathcal{O}$, the expectation value $\omega(A)$ is the average over the results of measurements of the observable $A$. Thus, a state of a system is a functional $\omega:\mathcal{O}\rightarrow \mathbb{R}$ that, of course, should satisfy  some extra mathematical conditions as explained below.

\section{Systems in Classical Mechanics}

\subsection{Observables in Classical Mechanics}

A typical system in Classical Mechanics\footnote{It is possible to generalize the discussion to more general symplectic or even Poisson manifolds but I will restrict the discussion to cotangent bundles.} is described in the cotangent bundle $T^*C$ of a smooth configuration space $C$; the space of all possible positions $q\in C$ that it can attain (possibly subject to external constraints). On the other hand, the  phase space $T^*C$  consists of  all possible values of position and momentum variables\footnote{I will use the Penrose abstract notation. Indices $a$, $b$, \dots refer to the manifold $C$ and $\alpha$, $\beta$, \dots to the manifold $T^*C$. This means, for instance, that given $q\in C$, the points of $T^*C$ are generically denoted by $(q,p_a)$, where $p_a\in T^*_qC$.} $(q,p_a)\in T^*C$. In this description the \textit{classical observables} belong to some class of functions on $T^*C$, for instance $\mathcal{O}_c=C^\infty(T^*C;\mathbb{R})\subset C^\infty(T^*C;\mathbb{C})\,.$ It is clear that $\mathcal{O}_c$ is a commutative and associative $*$-algebra of complex smooth functions whose elements satisfy the reality condition $A=A^*$, where the $*$ denotes the standard complex conjugation. As we will discuss later, this commutative structure will be responsible for the fact that there does not exist  an `uncertainty principle' in Classical Mechanics. There are other natural algebraic structures that can be defined on $\mathcal{O}_c$, some of them non-commutative. In particular, a very important one is the standard symplectic structure $\Omega_{\alpha\beta}$ on$T^*C$. It endows $\mathcal{O}_c$ with the structure of a Poisson $*$-algebra with Poisson bracket
$$
\{A,B\}:=\Omega^{\alpha\beta} (\mathrm{d} A)_\alpha (\mathrm{d} B)_\beta\in \mathcal{O}_c\,, \quad \forall A,B\in \mathcal{O}_c\,,
$$
where $\Omega^{\alpha\beta}$ is the inverse of $\Omega_{\alpha\beta}$, i.e. $\Omega^{\alpha \gamma}\Omega_{\gamma \beta}=\delta^\alpha\,_\beta$.
This Poisson structure is relevant in many respects. First, once the classical Hamiltonian of the system  $H\in \mathcal{O}_c$ is given, the dynamical evolution is defined through $\{\cdot,\cdot\}$ by
$$
\frac{d A}{d t}=\{A,H\}\,.
$$
Second,  as we will see, the Poisson bracket can be considered as the classical analogue of the quantum commutator and in this respect it represents a kind of shadow of the quantum world.

Two particular classes of observables are particularly important: the configuration and momentum observables. The class of \textit{configuration observables}  consists of  observables  $Q(f)\in \mathcal{O}_c$ of the form $Q(f)(q,p_a):=f(q)$,
where  $f\in C^\infty(C)$. The class of \textit{momentum observables} is parameterized by smooth vector fields on $C$, denoted generically by $v\in\mathfrak{X}^\infty(C)$, $q\mapsto (q,v^a(q))\in T_qC$. Its elements, $P(v):T^*C\rightarrow \mathbb{R}$, are defined using the fact that vector fields can be thought of as linear functions on the cotangent bundle  $P(v)(q,p_a):=p_a v^a(q)\,.$
The joint family of configuration and momentum observables is closed under Poisson brackets
$$
\{Q(f_1),Q(f_2)\}=0,\quad \{Q(f),P(v)\}=Q(\mathcal{L}_vf),\quad \{P(v_1),P(v_2)\}=-P(\mathcal{L}_{v_1}v_2)\,.
$$
Here $\mathcal{L}_v$ denotes the Lie derivative in the direction of $v$. This family contains many important observables such as the usual position, linear momentum, and angular momentum. For instance, if the configuration space is familiar Euclidean 3-space we can choose global Euclidean coordinates $x$, $y$, $z: C\rightarrow \mathbb{R}$ to build the standard position observables $X=Q(x)$, $Y=Q(y)$, and $Z=Q(z)$.
Then, the momentum observables
$P_X=P(\partial_x)$, $P_Y=P(\partial_y)$,  and $P_Z=P(\partial_z)$
are the components of the usual linear momentum. By considering the remaining three Killing vector fields of the Euclidean metric we can define the components of the familiar angular momentum $L_X=P(y\partial_z-z\partial_y)$, $L_Y=P(z\partial_x-x\partial_z)$, and $L_Z=P(x\partial_y-y\partial_x)$. The pairs of observables $(X,P_X)$, $(Y,P_Y)$, and $(Z,P_Z)$ are called \textit{canonical} and satisfy $\{X,P_X\}=\{Y,P_Y\}=\{Z,P_Z\}=1$.

\subsection{States in Classical Mechanics}

In Classical Mechanics it is always  assumed that canonical variables can be simultaneously measured with infinite  precision. This leads us to identify the points of the phase space with the \emph{pure states} of the system.
If  $z=(q,p_a)\in T^*C$ is determined with total precision, the expectation value of any observable $A\in\mathcal{O}_c$ is given by $
\omega_z(A)=A(q,p_a)\,.$
An experiment performed on a system described by a pure state will attain  maximal theoretical accuracy: If we define the variance $\Delta^2_\omega (A)$ of an observable $A$ relative to the state $\omega$  by
$$
\Delta^2_\omega (A) := \omega(A^2)-\omega(A)^2\,,
$$
it is clear that pure states satisfy  $\Delta_\omega(A)=0$ for all $A\in \mathcal{O}_c$.  In fact, they are characterized as the linear functional on $\mathcal{O}_c$ that satisfies $\omega(A_1A_2)=\omega(A_1)\omega(A_2)$ for all $A_1$, $A_2\in \mathcal{O}_c$. Pure states play a fundamental role in (non-statistical) Mechanics.

Although it is always assumed  that any system is in a pure state, we may be unable to determine it. For example, if we are dealing with a system with a very large number of particles, say $10^{23}$,  it is impossible in practice to determine all the positions and all the velocities of them. In these cases we are forced to work with an effective description of our system. Suppose, for instance, that we know that  a  system is in a state $\omega_{z_1}$ with probability $p$ and in a state $\omega_{z_2}$ with probability $1-p$. Then the effective state of the system is  is the \textit{mixture} of the states $\omega_{z_1}$ and $\omega_{z_2}$ defined by
$$
\omega(A)= p \,\omega_{z_1}(A)+(1-p)\,\omega_{z_2}(A)\,,\quad A\in\mathcal{O}_c.
$$
This is the usual situation in Classical Statistical Mechanics. In order to incorporate these situations to our description we are forced to identify the classical space of states of a system  $\mathcal{S}_c$ with the space of probability measures $\mu$ on $T^*C$. Given a probability measure $\mu$ we can define the state $\omega_\mu:\mathcal{O}_c\rightarrow \mathbb{R}$ as the linear functional
$$
\omega_\mu(A)=\int_{T^*C} A \, \mathrm{d}\mu\,.
$$
Pure states correspond to singular measures whose support is concentrated on some point of the phase space (i.e. Dirac deltas) and they cannot be written as a mixture of different states. Summarizing, the algebra of observables of Classical Mechanics can be realized as an algebra of \textit{random variables} on a probability space.

\section{Systems in Quantum Mechanics}
Classical Mechanics, whose geometric flavor makes it extremely beautiful, is a very natural way to model physical systems. However it suffers from one serious problem: Nature does not behave as  Classical Mechanics predicts. Here we will focus only on one point where Classical Mechanics does not predict the right experimental results: the Heisenberg uncertainty principle. The Heisenberg principle is summarized by the following, almost iconic, expression  $(\Delta_\omega X) (\Delta_\omega P_X) \geq \hbar/2$. What it actually states is that, irrespectively of the state $\omega$ of the system, the product of the standard deviations of two canonical variables is bounded from below by a non-zero constant. As we have discussed above this cannot be the case if Classical Mechanics is the right theory. In order to model Nature, we need to reconsider the set of properties  that the set of observables describing  a  physical system must satisfy.

\subsection{Observables in Quantum Mechanics}

Segal's postulates \cite{Segal1} try to encode the minimal set of properties that the class of observables $\mathcal{O}$ for \textit{any} physical theory (such as QM) should satisfy.  A \textit{Segal system} is a set $\mathcal{O}$ endowed with  the following structure: (S1) $\mathcal{O}$ is a linear space over $\mathbb{R}$. (S2) $(\mathcal{O},||\cdot||)$ is a real Banach space. (S3)  $A\mapsto A^2$ is a continuous function on $(\mathcal{O},||\cdot||)$. (S4)  $||A^2||=||A||^2\,,$  and $|| A^2-B^2|| \leq \max(||A^2||,||B^2||)$. All those postulates are easy to justify on physical grounds \cite{Segal1,Segal2}.
From an operational point of view, only bounded observables play a fundamental role. The norm of an observable is to be thought of as its maximum numerical value. Also, if $A_1$ and $A_2$ are bounded observables, it is possible to justify that  the linear combination $\lambda_1 A_1+\lambda_2A_2$ can be defined as an observable.

There are two disjoint classes of mathematical systems that satisfy Segal's postulates. On one hand we have \textit{special Segal systems}. For those systems there exists an associative $C^*$-algebra $\mathcal{A}$ with identity, $\textbf{1}\in\mathcal{A}$, such that  $\mathcal{O}=\{A\in \mathcal{A}\,|\, A=A^* \quad \textrm{(i.e. }A\textrm{ is self-adjoint)}\}$ and  $\mathcal{A}$ is generated by $\mathcal{O}$. There are also the so called \textit{exceptional Segal systems}  for which this is  not the case. Exceptional Segal systems are difficult to construct in practice and, so far, no one has been able to give an interesting physical application of them. Here we will assume that  physical systems are special Segal systems. Hence, from a mathematical point of view, a  physical system is defined in terms of a  $C^*$-algebra $\mathcal{A}$ with unit $\mathbf{1}$.

\subsection{States in Quantum Mechanics}

The  states $\mathcal{S}$ of a Segal system $\mathcal{A}$, or simply the states over $\mathcal{A}$, are normalized positive linear functionals on $\mathcal{A}$ that separate points. Recall that a linear functional $\omega$ is called positive in $\mathcal{A}$ if $\omega(A^*A)\geq 0\,,$  $\forall A\in \mathcal{A}$. All positive functionals are continuous on $(\mathcal{A},||\cdot||)$. We will say that $\omega$ is normalized if $\omega(\mathbf{1})=1$. The normalization condition is necessary to give an statistical interpretation to the states. Finally, a class of linear functionals $\mathcal{S}$ is said to separate observables if for any pair $A_1\neq A_2$ there exists  $\omega\in \mathcal{S}$ such that $\omega(A_1)\neq \omega(A_2)$.  Notice that the set of states over $\mathcal{A}$ is a convex subset of $\mathcal{A}^*$ (the topological dual of $\mathcal{A}$). In this algebraic scheme a state is called \textit{pure} if it cannot be written as a nontrivial convex combination of other states. Otherwise the state is called non-pure or mixture.

\paragraph{Simultaneous observability}
An observable $A$ is said to have a definite value in a state $\omega$ if $\Delta^2_\omega(A):=\omega(A^2)-\omega(A)^2=0$. Observables in a class are called simultaneously observable if there exists a sufficient large number of states in which they simultaneously take definite values. Explicitly, a collection $\mathfrak{C}$ of observables is \textit{simultaneously observable} if the system $\mathcal{A}(\mathfrak{C})$ generated by $\mathfrak{C}$ has a set of states, that separates points of $\mathcal{A}(\mathfrak{C})$, such that every observable in $\mathcal{A}(\mathfrak{C})$ has a definite value in each of them. The class of simultaneous observables can be characterized by the following theorem

\noindent\textbf{Theorem.} $\mathfrak{C}$  \textit{is simultaneously observable if and only if it is commutative.}

The abelian abelian $C^*$-algebras were characterized through the works of Gelfand and Naimark.
An abelian $C^*$-algebra $\mathcal{A}$ with identity is isometrically isomorphic to the $C^*$-algebra of continuous functions on a compact Hausdorff topological space (whose topology is induced by the weak $*$ topology)  called the Gelfand spectrum of $\mathcal{A}$, $sp(\mathcal{A})$.
If $\mathcal{A}$ is abelian with identity there exists an isomorphism  $\mathcal{A} \rightarrow  C(sp(\mathcal{A}))$, $A\mapsto f_A$. The Riesz-Markov theorem ensures that associated to every state $\omega$  there is a measure $\mu$ on $sp(\mathcal{A})$  such that
$$
\omega(A)=\int_{sp(\mathcal{A})} f_A d\mu\,.
$$
Then, if $\mathfrak{C}$ is a simultaneously observable class it is isomorphic to the system of all real valued continuous functions on the compact Hausdorff space $sp(\mathcal{A}(\mathfrak{C}))$. In this case, the situation for $\mathfrak{C}$ is exactly the same as in Classical Mechanics in the sense that the algebra of observables of an abelian $C^*$-algebra can be realized as an algebra of \textit{random variables} on a probability space. Then, in order to incorporate the Heisenberg uncertainty relations, we are forced to consider  non-abelian $C^*$-algebras to describe QM.

\paragraph{Heisenberg uncertainty relations}

The Heisenberg uncertainty relations are tied to the non-commutative character of the $C^*$-algebra of observables through the following (easy to prove) theorem.

\noindent\textbf{Theorem (Heisenberg uncertainty relations)}
\textit{Given two observables} $A_1,A_2\in\mathcal{A}$ \textit{ and a state} $\omega$ we have that $
\Delta_\omega(A_1)\cdot \Delta_\omega(A_2)\geq |\omega([A_1,A_2])|/2$, \textit{where the commutator} $[A_1,A_2]$ \textit{is defined by}  $[A_1,A_2]:=A_1A_2-A_2A_1\,.$

\paragraph{Statistical interpretation of Quantum Mechanics}
The interpretation of a system described by a non-commutative algebra of observables is intrinsically statistical. To see this consider a normal element $A\in \mathcal{A}$  (i.e. $AA^*=A^*A$). The $C^*$-algebra $\mathcal{A}(A)$ generated by  $\mathbf{1}$, $A$, and $A^*$ is abelian and its spectrum satisfies
$$sp(\mathcal{A}(A))=\sigma(A)=\{\lambda\in \mathbb{C}\,|\, \lambda \mathbf{1}-A \textrm{ does not have a two-sided inverse}\}.$$
Then, given any state $\omega$ there exists a $\mu_{\omega,A}$ on $sp(\mathcal{A}(A))$ such that
$$
\omega(B)=\int_{\sigma(A)} f_B(\lambda) \mathrm{d}\mu_{\omega,A} (\lambda)\,, \quad \forall B\in \mathcal{A}(A)\,.
$$
In particular
\begin{equation}
 \omega(A)=\displaystyle \int_{\sigma(A)} \lambda \mathrm{d}\mu_{\omega,A} (\lambda)\,.\label{mu_A}
\end{equation}
Recalling  that $\omega(A)$ must be understood  as the expectation value of $A$ on $\omega$, the interpretation of equation (\ref{mu_A}) is clear. If $A$ is an observable ($A=A^*$), the possible values that $A$ can take in any experiment belong to its spectrum $\sigma(A)$. The probability that, when the state is $\omega$, the observable $A$ takes values on certain subset of $\sigma(A)$ is given in terms of $\mu_{\omega,A}$. It is important to notice that if $\mathcal{A}$ is not abelian the probability measures  $\mu_{\omega,A}$ associated with pure states are not Dirac-like measures. Then, in contrast with Classical Mechanics, the statistical interpretation of QM cannot be avoided by restricting the states to be pure.

\paragraph{Representations of $C^*$-algebras} It can be shown that any $C^*$-algebra is isomorphic to an algebra of bounded operators in a Hilbert space. The theory of $C^*$-algebra representations  provides concrete realizations of the abstract algebra of observables of a physical system and also allows the implementation of the superposition principle of wave mechanics. As we will see, given a representation of $\mathcal{A}$ on a Hilbert space $\mathfrak{H}$, the observables can be thought of as bounded self-adjoint operators on $\mathfrak{H}$ and the unit vectors $\psi\in \mathfrak{H}$ define pure states.

A representation $(\mathcal{A},\varrho,\mathfrak{H})$, or simply $\varrho$, of a $C^*$-algebra $\mathcal{A}$ in a Hilbert space $\mathfrak{H}$ is a $*$-homomorphism $\varrho$ of $\mathcal{A}$ into the $C^*$-algebra $\mathcal{B}(\mathfrak{H})$ of bounded linear operators in $\mathfrak{H}$. We will focus our attention on \textit{faithful}  and \textit{irreducible} representations i.e. such that $ker(\varrho)=\{0\}$ and $\{0\}$ and $\mathfrak{H}$ are the only closed subspaces invariant under $\varrho(\mathcal{A})$ representations.

One of the most important results concerning the theory of representations of $C^*$-algebras is the GNS theorem.

\noindent\textbf{Theorem (Gelfand-Naimark-Segal)}\textit{ Given a $C^*$-algebra $\mathcal{A}$ with identity and a state $\omega$, there is a Hilbert space $\mathfrak{H}_\omega$ and a representation $\varrho_\omega:\mathcal{A}\rightarrow \mathcal{B}(\mathfrak{H})$ such that}

\begin{itemize}
\item \textit{$\mathfrak{H}_\omega$ contains a cyclic vector $\psi_\omega$ (i.e. $\overline{\varrho(\mathcal{A})\psi_\omega}=\mathfrak{H}_\omega$).}

\item \textit{$\omega(A)=\langle \psi_\omega\,|\,\varrho_\omega(A)\psi_\omega\rangle$ for all $A\in\mathcal{A}$.}

\item \textit{Every other representation $\varrho$ in a Hilbert space $\mathfrak{H}$ with a cyclic vector $\psi$ such that $\omega(A)=\langle \psi\,|\,\varrho(A)\psi\rangle \quad\forall A\in\mathcal{ A}$
is unitarily equivalent to $\varrho_\omega$, i.e. there exists an isometry $U:\mathfrak{H}\rightarrow \mathfrak{H}_\omega$ such that
$U\psi=\psi_\omega\,,\quad U\varrho(A)U^{-1}=\varrho_\omega(A)$ for all $A\in \mathcal{A}$.}
\end{itemize}

Notice that if we are given a representation $(\mathcal{A},\varrho,\mathfrak{H})$, the \textit{unit} vectors $\psi\in \mathfrak{H}$ define states $\omega_\psi$ on $\mathcal{A}$ through the formula $\omega_\psi(A):=\langle \psi \,|\, \varrho(A)\psi\rangle\,.$ These states are called \textit{state vectors} of the representation. The converse is also true by the GNS theorem.  However it is possible to show that if a state is not pure the representation $(\mathfrak{H}_\omega,\varrho_\omega,\psi_\omega)$ is reducible.

\noindent\textbf{Theorem.} \textit{Let $\omega$ a state over the $C^*$-algebra $\mathcal{A}$ and $(\mathfrak{H}_\omega,\varrho_\omega,\psi_\omega)$ the associated cyclic representation. Then $(\mathfrak{H}_\omega,\varrho_\omega,\psi_\omega)$ is irreducible iff $\omega$ is pure.}

It is also possible to build non-pure states that represent our ignorance about the pure state of a system.  Given a positive trace class operator $b$ on $\mathfrak{H}$, with trace equal to one, the formula $\omega_b(A):=\mathrm{tr}(b\varrho(A))$
defines a state over $\mathcal{A}$. These states are called \textit{density matrices}. Notice that $\omega_b$  is of the form
$\omega_b(A)=\sum_i p_i \langle\psi_i\,|\, \varrho(A)\psi_i\rangle\,,$ where $p_i\geq 0$,  $\sum_ip_i=1$, and $\langle\psi_i\,|\, \psi_i\rangle =1$. Then the state $\omega_b$ is a pure state if and only if $b$ is a one dimensional projection.

\section{The quantum particle and the Weyl algebra}
Although theoretically the operational approach doesn't need anything more than experimental inputs, in practice the usual way to construct observables  takes  the classical description of the system as the starting point and follows some kind of `quantization procedure'. These quantization rules,  usually known as Dirac's quantum conditions, assume that there exists a map
$
\hat{\phantom{f}}: \tilde{\mathcal{O}}_c\subset\mathcal{O}_c \rightarrow \mathcal{O}
$
from a subset $\tilde{\mathcal{O}}_c$ of the classical observables into the quantum ones such that  $A\mapsto \hat{A}$ is linear over $\mathbb{R}$; if $A$ is a constant function, then $\hat{A}$ is the corresponding multiplication operator,
and if $\{A_1,A_2\}=A_3$ then $[\hat{A}_1,\hat{A}_2]=-i\hbar \hat{A}_3$. Notice that the last  rule, that relates the  classical Poisson bracket with the quantum commutator, assumes that $\tilde{\mathcal{O}}_c$ is closed under $\{\cdot,\cdot\}$.

Let us illustrate these quantization rules in the simplest situation: the spinless point particle.

\paragraph{Heisenberg algebra}
The simplest quantum system is the one-dimensional particle.  Let us try to construct its  $C^*$-algebra.  The basic classical observables are position $X$ and momentum $P=P_X$ so, naively, one can try to consider the algebra of observables generated by $X$ and $P$ which satisfy the Heisenberg commutation relations $[P,X]=i$, $[X,X]=0$, and  $[P,P]=0$. Here and in the following we choose units such that $\hbar=1$.

The problem with this approach is that the Heisenberg algebra \textit{does not} fall into the Segal scheme. In particular $X$ and $P$ cannot be self-adjoint elements of any $C^*$-algebra because $||X||$ and $||P||$ cannot both be finite. In particular $[P,X^n]=-inX^{n-1}$ implies that $||X|| ||P||\geq n/2\,,\quad \forall n\in \mathbb{N}$.  Then $X$ and $P$ are not observables in the operational sense.

\paragraph{Weyl algebra} The solution to the problems associated with the Heisenberg algebra in the operational approach were solved by Weyl by considering the polynomial algebra generated by the classical observables $
U(\alpha)=\exp(i\alpha X)$ and $V(\beta)=\exp(i\beta P)$ and following the quantization rules stated above. Explicitly,  the \textit{Weyl algebra} is  the algebra  generated (through complex linear combinations
and products) by the elements $\{U(\alpha), V(\beta)\,|\, \alpha,
\beta\in \mathbb{R}\}$ constrained to satisfy the following conditions: $U(0)=V(0)=\mathbf{1}$; $U(\alpha)^*=U(-\alpha)\,,\quad V(\beta)^*=V(-\beta)$;  $U(\alpha)^*U(\alpha)=U(\alpha)U(\alpha)^*=\mathbf{1}=V(\beta)^*V(\beta)=V(\beta)V(\beta)^*$;  $U(\alpha_1)U(\alpha_2)=U(\alpha_1+\alpha_2)\,,$
$ V(\beta_1)V(\beta_2)=V(\beta_1+\beta_2)\,,$ $U(\alpha)V(\beta)=V(\beta)U(\alpha)\exp(-i\alpha\beta)$; and $||U(\alpha)||=||V(\beta)||=||U(\alpha)V(\beta)||=1$.  Then the \textit{Weyl $C^*$-algebra} $\mathcal{A}_{\mathrm{Weyl}}$ is the $||\cdot||$-completion of the Weyl algebra. In QM it is assumed that the quantum particle is the physical system characterized by $\mathcal{A}_{\textrm{Weyl}}$.

\paragraph{Representations of the Weyl algebra} The classification of  the representations of $\mathcal{A}_{\textrm{Weyl}}$ is solved by the following uniqueness theorem due to von Neumann.

\noindent{\textbf{Theorem (von Neumann)}} \textit{All the regular irreducible representations of} $\mathcal{A}_{\textrm{Weyl}}$ \textit{in separable  Hilbert spaces are unitarily equivalent.}

Here  `regular'  means that the one dimensional families of unitary operators $\varrho(U(\alpha))$ and $\varrho(V(\beta))$ must be strongly continuous in $\alpha$ and $\beta$ respectively.

\paragraph{The Schr\"odinger representation} The Schr\"odinger representation $(\mathcal{A}_{\textrm{Weyl}},\varrho,\mathfrak{H})$ is a regular irreducible representation of $\mathcal{A}_{\textrm{Weyl}}$ in the  separable Hilbert space $\mathfrak{H}=L^2(\mathbb{R})$. Denoting $\hat{U}(\alpha):=\varrho(U(\alpha))$ and  $\hat{V}(\beta):=\varrho(V(\beta))$, the Schr\"odinger representation is defined by $
\big(\hat{U}(\alpha)\psi\big)(x):=e^{i\alpha x}\psi(x)$ and $\big(\hat{V}(\beta)\psi\big)(x):=\psi(x+\beta)$ for any  $\psi\in \mathfrak{H}$.
By using Stone's theorem, the Schr\"odinger representation provides also a representation of the Heisenberg algebra $(\hat{X}\psi)(x)=x\psi(x)$,  $(\hat{P}\psi)(x)=i\psi'(x)$. The position and momentum operators are unbounded and, hence, only densely defined. The position operator  acts as a multiplicative operator whereas the momentum operator acts as a derivative operator over the vector states of the Schr\"odinger representation.

\section{algebraic dynamics}

We are now interested in describing the relations between measurements at different times for non-dissipative systems.  In this case it it plausible to demand the following:
\begin{itemize}

\item The relations between the measurements at times $t_1$ and $t_2$ depend only on the difference $t_2-t_1$.

\item If an observable $A$ is defined by some experimental device at a given time, say $t=0$, the same type of measurements performed at time $t$ defines an observable $A_t$.

\item The algebra $\mathcal{A}$ generated by the observables is the same at any time.

\item The time translation $A\mapsto \alpha_t(A)=A_t$ is a $*$-automorphism (preserves the algebraic properties).

\item For any state $\omega$ and any observable $A$, the real function $t\mapsto \omega(\alpha_t(A))$ is a continuous function.

\end{itemize}
An \textit{algebraic dynamical system} is a triplet $(\mathcal{A},\mathbb{R},\alpha)$
where $\mathcal{A}$ is a $C^*$-algebra and  $\alpha_t$ are, for each $t\in \mathbb{R}$,  automorphisms of $\mathcal{A}$ that satisfy the following conditions  $\alpha_0=\mathrm{id}$, $\alpha_{t_1}\circ \alpha_{t_2}=\alpha_{t_1+t_2}$, and
 $\alpha$ is weakly continuous i.e. $t\mapsto \omega(\alpha_t(A))$ is continuous for all $\omega$  and  $A$.

\paragraph{Dynamics and representations}
Given a  representation $(\mathcal{A},\varrho,\mathcal{H})$  of the $C^*$-algebra of observables we will say that $\varrho$ is stable under the evolution defined by $\alpha_t$ if  $\varrho$ and $\varrho\circ \alpha_t$ are unitarily equivalent. In other words, there is a unitary operator $ \hat{U}(t):\mathfrak{H}\rightarrow \mathfrak{H}$ such that $\varrho(\alpha_t(A))=\hat{U}^{-1}(t)\varrho(A)\hat{U}(t)$ for every observable $A$.
The weak continuity of $\alpha_t$ implies the weak continuity of $\hat{U}(t)$. Applying Stone's theorem we can write  $\hat{U}(t)=\exp(-it\hat{H})$
for some self-adjoint operator $\hat{H}$, with dense domain $\mathcal{D}(\hat{H})\subset \mathfrak{H}$. The self-adjoint operator $\hat{H}$ is called the quantum Hamiltonian (in the representation $\varrho$) of the dynamical system. Notice that the Hamiltonian \textit{is} a representation dependent concept and, in general, is an unbounded operator that does not belong to the $C^*$-algebra generated by the physical observables.

\paragraph{Heisenberg equation}
The Heisenberg equation is the time evolution differential equation for observables in a given stable representation.
Let us consider at $t=0$ an observable $A_0$ represented by $\hat{A}_0:=\varrho(A_0)$. The observable defined at time $t$ in terms of the same class of physical measurements than $\hat{A}_0$ is  $\hat{A}(t):=U(t)^{-1}\hat{A}_0U(t)$. Differentiating  $\hat{A}_t$ with respect to $t$ we get
$$
\frac{\mathrm{d}\hat{A}}{\mathrm{d}t}=i[\hat{H},\hat{A}]\,,\quad  A(0)=A_0\in \mathcal{B}(\mathfrak{H})\,.
$$
The equation above is called the Heisenberg equation and describes the time evolution of an observable whose initial value, at $t=0$, is $\hat{A}_0$.

\paragraph{The Schr\"odinger equation}
The evolution in a given stable representation can be equivalently formulated in terms of states.
The Schr\"odinger equation is the time evolution differential equation for pure states in a given stable representation. Let $\psi_0\in \mathcal{D}(\hat{H})$ be a vector state of a stable representation with unitary evolution operator given by $\hat{U}(t)=\exp(-it\hat{H})$ and let us denote by $\omega_0$ the pure state defined by $\psi_0$. Then
$$
\omega_0(\alpha_t(A))=\langle \psi_0 \,|\, \hat{U}(t)^{-1}\varrho(A) \hat{U}(t) \psi_0 \rangle =\langle \hat{U}(t) \psi_0 \,|\, \varrho(A) \hat{U}(t) \psi_0 \rangle =\omega_t(A)\,,
$$
where $\omega_t$ is the pure state defined by the unit vector $\psi(t):=\hat{U}(t)\psi_0\in \mathfrak{H}$.
By  differentiating   $\psi(t)$ with respect to $t$  we get the so called \textit{Schr\"odinger equation}
$$
i\frac{\mathrm{d}\psi}{\mathrm{d}t}=\hat{H}\psi\,,\quad \psi(0)=\psi_0\in \mathcal{D}(\hat{H})\,,
$$
that describes the time evolution of the initial vector state $\psi_0$.

\paragraph{The quantum particle in a potential}
If we consider an explicit quantum Hamiltonian $\hat{H}$ in the Schr\"odinger representation, the Heisenberg equations for the position and momentum operators $\dot{\hat{X}}=i[\hat{H},\hat{X}]$, $\dot{\hat{P}}=i[\hat{H},\hat{P}]$
are the analogs of the classical Hamilton equations.  If we particularize these equations for the class of `formal'  Hamiltonians of the form   $H(X,P)=\frac{1}{2}P^2+V(X)$ we get the Heisenberg equations for a particle in a potential: $\dot{\hat{X}}=\hat{P}$,
$\dot{\hat{P}}=V'(\hat{X})$. From the mathematical point of view we need to show that the  `quantum Hamiltonian' $\hat{H}=\frac{\hat{P}^2}{2}+V(\hat{X})$ is actually a well defined self-adjoint operator in the Schr\"odinger  representation. It is not difficult to show that, in general,  a expression of the form $\frac{1}{2}(\hat{P}_X^2+\hat{P}_Y^2+\hat{P}_Z^2)+V(\hat{X},\hat{Y},\hat{Z})$ defines a symmetric operator on $L^2(\mathbb{R}^3)$. However, this is not enough to have a well defined quantum Hamiltonian  because  symmetric operators may have several or even no self-adjoint extensions. This problem can be handled in many practical situations by using some of the theorems due to Kato.

\noindent\noindent \textbf{A taste of Kato's theorems:} \textit{For a potential $V$ in a ``Kato class'', the Cauchy problem for the Schr\"odinger equation}
$$
 i\frac{\partial \psi}{\partial t}(\vec{x},t)=-\frac{1}{2}\Delta \psi(\vec{x},t) +V(\vec{x})\psi(\vec{x},t)\,,\quad \psi(\vec{x},0)=\psi_0(\vec{x})\in \mathcal{D}(\Delta)\,,
$$
\textit{is well posed and the corresponding Cauchy problem has a unique solution global in time. Here $\Delta$ denotes the Laplacian of the 3-dimensional Euclidean space.}

The Coulomb potential $V(\vec{x})=-1/r=-1/\sqrt{x^2+y^2+z^2}$ belongs to the Kato class $V\in L^2_{\mathrm{loc}}(\mathbb{R}^3)$, $V\geq 0$. Due to this fact, most of the problems of atomic and nuclear physics have a unique solution, global in time. In particular the hydrogen atom can be described by a self-adjoint Hamiltonian of the form $H_0=-\Delta/2-1/r$  and it is possible to show that the negative spectrum of $H_0$ is discrete, in agreement with the experimental results.

\bibliographystyle{aipproc}   

\end{document}